\documentclass[aps,prl,superscriptaddress,twocolumn]{revtex4}
\usepackage{bm,graphicx,graphics,amsmath,amssymb,bm,epsfig,color,pifont}
\usepackage{euscript,tabularx,textcomp}
\usepackage[english]{babel}

\begin{document}
 
\title{Magnetization and EPR studies of the single molecule magnet Ni$_4$\\
  with integrated sensors}

\author{G. de Loubens} \affiliation{Department of Physics, New York
  University, 4 Washington Place, New York, New York 10003, USA}
\author{G. D. Chaves-O'Flynn} \affiliation{Department of Physics, New York
  University, 4 Washington Place, New York, New York 10003, USA}
\author{A. D. Kent} \affiliation{Department of Physics, New York
  University, 4 Washington Place, New York, New York 10003, USA}

\author{C. Ramsey}\affiliation{Physics Department, University of
  Central Florida, 4000 Central Florida Boulevard, Orlando, Florida
  32816-2385, USA}
\author{E. del Barco}\affiliation{Physics Department, University of
  Central Florida, 4000 Central Florida Boulevard, Orlando, Florida
  32816-2385, USA}

\author{C. Beedle}\affiliation{Department of Chemistry and
  Biochemistry, University of California San Diego, La Jolla,
  California 92093-0358, USA}
\author{D. N. Hendrickson}\affiliation{Department of Chemistry and
  Biochemistry, University of California San Diego, La Jolla,
  California 92093-0358, USA}

\date{\today}

\begin{abstract}
  Integrated magnetic sensors that allow simultaneous EPR and
  magnetization measurements have been developed to study single
  molecule magnets. A high frequency microstrip resonator has been
  integrated with a micro-Hall effect magnetometer.  EPR spectroscopy
  is used to determine the energy splitting between the low lying
  spin-states of a Ni$_4$ single crystal, with an $S=4$ ground state,
  as a function of applied fields, both longitudinal and transverse to
  the easy axis at 0.4~K. Concurrent magnetization measurements show
  changes in spin-population associated with microwave absorption.
  Such studies enable determination of the energy relaxation time of
  the spin system.
\end{abstract}


\maketitle

An understanding of decoherence and energy relaxation mechanisms in
single molecule magnets (SMMs) is both of fundamental interest
\cite{chudnovsky05} and important for the use of SMMs in quantum
computing \cite{leuenberger01}. Quantum tunneling of magnetization
(QTM) has been widely studied in SMMs
\cite{friedman96,thomas96,wernsdorfer99}. A recent focus is on
coherent QTM in which the tunneling rates may be faster than the rate
of decoherence \cite{hill03,barco04b}. Experiments on SMMs
\cite{barco04b} and doped antiferromagnetic rings
\cite{wernsdorfer05b, ardavan06} have been reported. However, the
basic relaxation mechanisms in SMMs are still under active
investigation both experimentally and theoretically
\cite{chudnovsky02}.

We have developed sensors allowing simultaneous EPR and magnetization
measurements at low temperatures (typically below 1~K) to study SMMs.
A microstrip resonator \cite{gupta96} with resonance frequency between
25 and 30~GHz has been integrated on a chip with a micro-Hall effect
magnetometer \cite{kent94}. The high filling factor in such resonators
allows measurement of photon absorption in very small crystals as well
as the application of large microwave magnetic fields, needed for Rabi
experiments. The fast response of the Hall sensor ($>1$~MHz) also
enables time-resolved measurements of the magnetization

In this paper we show EPR spectroscopy of the two lowest energy levels
associated with high spin states ($S=4$) of a Ni$_4$ single crystal.
We present simultaneous measurements of associated photon-induced
changes in the magnetization. This represents an important advance
from the experiments in refs.~\cite{barco04b,wernsdorfer05b}, in which
only photon induced magnetization changes were measured, and EPR
studies were not possible. In particular, it enables a direct
determination of the energy relaxation time of the spin system.

\begin{figure}
  \includegraphics[width=6cm]{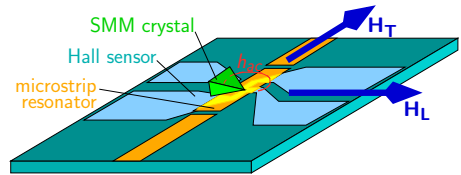}
  \caption{Schematic of the sensor developed for this study.}
  \label{setup}
\end{figure}

A schematic of our integrated sensor is presented in Fig.~\ref{setup}.
The Hall sensor is fabricated from a GaAs/AlGaAs heterostructure 2D
electron gas to form a cross of width $50~\mu$m, chosen to optimize the
coupling with a SMM crystal with lateral dimensions of about $100~\mu$m.
Our magnetometer has a Hall coefficient of $1780~\Omega/$T and a noise
level of $2~\mu$T$/\sqrt{Hz}$ with a current of $10~\mu$A. This permits
detection of changes in the magnetization smaller than $5\times10^{-4}$ of
the saturation magnetization of our SMM crystals ($\approx5\times10^5\mu$m$^3$),
using a lock-in detection bandwith of 300~ms. The microstrip
resonator, a 250~nm thick gold line, is evaporated on top of the Hall
sensor and capacitively coupled to two feedlines, so that it can work
in transmission mode. It is designed to be matched to 50~$\Omega$ and its
fundamental resonance frequency ranges between 25 and 30~GHz
($\approx0.4$~mm width and 1.5~mm length). The MW field, $h_{ac}$, is
maximum and uniform above its center, aligned with the Hall cross. The
strength of $h_{ac}$ at the sample location was calibrated for each
integrated sensor used: the circularly polarized saturation amplitude
of DPPH ($h_{ac}=0.9\pm0.05$~Oe) corresponds to an incident power
$P_{in}$ of about 50~mW. This value depends on the quality factor of
the resonator, found to be about 20.

The sensor is in a $^3$He cryostat with a base temperature of 0.35~K.
The two ports of the microstrip are connected through coaxial lines to
an Agilent PNA 50~GHz vector network analyzer with copper and
stainless steel sections to minimize thermal and MW losses (12 to
14~dB in our frequency range at low temperature). The PNA is used as a
MW source and allows transmission measurement. Furthermore, it is
possible to gate the source using a pulse pattern generator (no pulsed
experiments will be presented in this paper). A high field
superconducting vector magnet is used to apply DC magnetic fields in
arbitrary directions with respect to the axes of the crystal.

\begin{figure}
  \includegraphics[width=6cm]{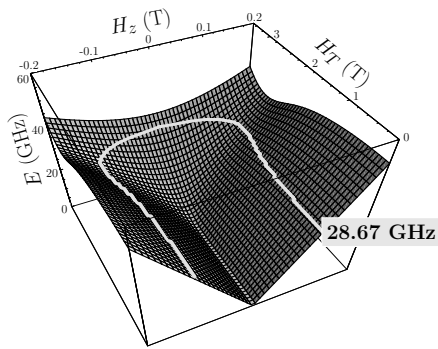}
  \caption{Energy separation $E$ between the two lowest energy states
    $|A\rangle$ and $|S\rangle$ versus $H_z$ and $H_T$. The locus of the resonance
    observed experimentally is a cut at constant energy, shown by the
    thick line.}
  \label{3D}
\end{figure}

To demonstrate the capabilities of our integrated sensor, we have
studied [Ni(hmp)(dmb)Cl]$_4$, henceforth referred to as Ni$_4$.  This
is a particularly clean SMM with no solvate molecules present in its
crystal phase and only 1\% (natural abundance) of nuclear spins on the
transition metal sites \cite{yang03}. This results in narrower EPR
peaks than in many SMMs \cite{edwards03}. The spin Hamiltonian of
Ni$_4$ is to first approximation:
\begin{equation}
  \mathcal{H}=-DS_z^2-\mu_B\vec{H}.\hat{g}.\vec{S}+\mathcal{H}',
\end{equation}
where the first term is the uniaxial anisotropy, the second the Zeeman
energy, and the last one includes higher order anisotropy terms
(\emph{i.e.}, $-BS_z^4+C(S_+^4+S_-^4)$). The $S=4$ ground state of the
molecule at low temperature is the consequence of ferromagnetic
exchange interactions between the four Ni$^{\text{II}}$ ($S=1$) ions.
The uniaxial anisotropy leads to a large energy barrier $DS^2\approx12$~K
(anisotropy field $H_a=2DS/(g\mu_B)=4.5$~T) to magnetization reversal
between states of projection $S_z=\pm4$ along the easy axis of the
molecules ($z$). The two lowest energy states are the symmetric
($|S\rangle$) and antisymmetric ($|A\rangle$) linear combinations of $|up\rangle$ and
$|down\rangle$ states. In the presence of a transverse field $H_T$, the
latter are tilted away from the easy axis and have opposite $z$
projections \cite{barco04b}. There is an energy separation between
$|A\rangle$ and $|S\rangle$, known as the tunnel splitting $\Delta$, which, in the
field range in these experiments, increases with $H_T$ when $H_z$
equals zero. This energy separation also increases with $H_z$ due to
the Zeeman term. Thus, when a sample is irradiated at constant
frequency $f$ with a MW field $h_{ac}$ aligned with $z$, it is
possible to induce transitions between $|A\rangle$ and $|S\rangle$ and to map out
the constant energy splitting ($=hf$) locus in the ($H_z$, $H_T$)
phase space (\emph{cf.}  Fig.~\ref{3D}). These measurements enable
determination of the spin-Hamiltonian parameters and set a lower bound
of the decoherence time, as shown below. The transition rate is given
by the Fermi golden rule:
\begin{equation}
  \Gamma=\frac{\pi}{2} \left ( \frac{g\mu_B}{\hbar}h_{ac} \right ) ^2|\langle S|S_z|A\rangle|^2f(\omega)
  \label{rate}
\end{equation}
where $\langle S|S_z|A\rangle$ is the matrix element coupling the
states and $f(\omega)$ is a lorentzian of linewidth $1/T_2$.

In steady state, the absorbed power is simply $P_{abs}=\Gamma hf n_d$,
where $n_d$ is the difference in population between states $|A\rangle$ and
$|S\rangle$. When these states have different projections on the $z$-axis,
the photon-induced transitions lead to a change in magnetization $\Delta
M$. If these projections on the $z$-axis are opposite, one can deduce
$\Delta M$ from energy conservation. It is simply linked to the absorbed
power and to the energy relaxation time $T_1$ \cite{bloembergen54}:
\begin{equation}
  T_1P_{abs}=\frac{hf\Delta MN_0}{2M_{eq}},
  \label{relax}
\end{equation}
where $N_0$ and $M_{eq}$ are the difference in population between
$|A\rangle$ and $|S\rangle$ and the magnetization in the absence of MW,
respectively.  Thus, simultaneous measurements of $P_{abs}$ and $\Delta M$
yield $T_1$. We emphasize that this latter may be different from the
spin-phonon relaxation time, as in the case of the phonon bottleneck
\cite{abragam70,garanin06}.

At the temperatures used in our experiments only the two lowest lying
spin-levels are thermally populated. Moreover, for the transverse
fields used, the thermal energy is smaller than the photon energy,
$k_BT<hf$. A pyramidal Ni$_4$ crystal is placed with one of its faces
parallel to the plane of the sensor, in the middle of the microstrip
resonator, and oriented so that $h_{ac}$ is aligned with $z$, along
the axis of the pyramid. The Hall device responds to the average
magnetic field perpendicular to the plane of the sensor, which for the
sample shape and placement, is mainly due to the $z$-component of the
sample magnetization. Because of this pyramidal shape, the easy axis
is not aligned ($\alpha=20$\textdegree) with $H_L$, which lies in the sensor plane.
There is also a second misalignment $\theta<5$\textdegree{} between $H_L$ and the
projection of the $z$ axis on the sensor plane. The longitudinal field
felt by the molecules is thus $H_z=H_L \cos \alpha \cos \theta +H_T \cos \alpha \sin
\theta$, whereas the transverse field is to a good approximation $H_T$.
Note that the EPR data are plotted versus $H_z$.

\begin{figure}
  \includegraphics[width=8cm]{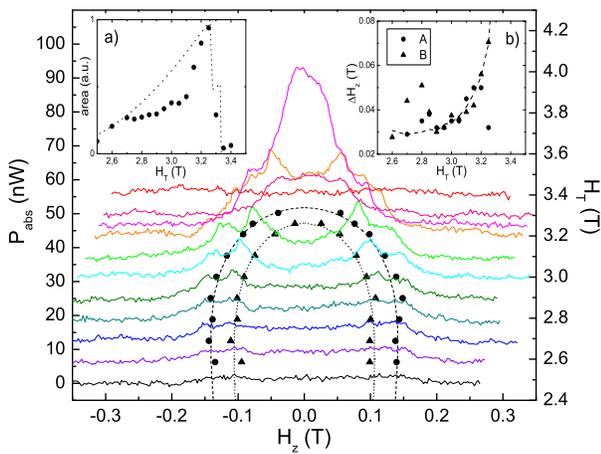}

  \caption{EPR absorption signal vs. longitudinal field $H_z$ for
    several transverse fields $H_T$. $f=28.67$~GHz, $P_{in}=2.5~\mu$W
    and $T=0.4$~K. The circles and triangles denote the position of
    resonances. The dashed lines are the result of a direct
    diagonalization of $\mathcal{H}$.  Insets: (a) integrated area
    under the EPR signal vs. $H_T$ (b) linewidth of peaks A and B vs.
    $H_T$.}
  \label{EPR}
\end{figure}

Fig.~\ref{EPR} presents the EPR signal measured at $f=28.67$~GHz for
several values of $H_T$ as $H_z$ is swept at 0.1~T$/$min. The power
transmitted through the resonator is monitored in cw mode. The low
power level used for this study induces a small increase of the sample
temperature to 0.4~K. Two symmetric groups of peaks with respect to
$H_z=0$~T are observed. They correspond to transitions between $|A\rangle$
and $|S\rangle$. A multiple-lorentzian fit leads to two main peaks, A and B,
assigned to two species of Ni$_4$ molecules present in the crystal,
associated with different molecular environments \cite{edwards03}. As
$H_T$ is increased, the amplitude of the peaks increases and the
resonance occurs at smaller $H_z$. As expected the EPR signal
disappears at large $H_T$ as $\Delta$ becomes larger than $hf$.  A direct
diagonalization of $\mathcal{H}$ yields the dependence of the energy
splitting on $H_z$ and $H_T$. The dashed line is a fit of the peak A
data (circles) using $D_A = 0.757$~K, $B = 7.9~10^{-3}$~K, $C =
3.25~10^{-5}$~K, $g_z=2.3$, $g_x=g_y=2.23$. These parameters are in
good agreement with high frequency cavity EPR measurements
\cite{edwards03}. The dotted line is the best fit for peak B
(triangles), obtained for $D_B = 0.785$~K, $B = 9.9~10^{-3}$~K, $C =
3.25~10^{-5}$~K, $g_z=3.02$, $g_x=g_y=2.39$. An extremely high value
(unphysical) of $g_z$ is needed to fit the data points, which is not
understood at present.

The curvature of the energy splitting as the longitudinal field goes
to zero is evidence of level repulsion, as already noted in
ref.~\cite{barco04b}. EPR has the advantage that one can meausre the
splitting at small $H_z$, where no or only small changes in the
magnetization are expected. The inset (a) of Fig.~\ref{EPR} shows the
area under the experimental absorption curves versus $H_T$. The dashed
line is the prediction from the direct diagonalization of
$\mathcal{H}$ and Eq.~\ref{rate}. The calculation shows some
discrepancy with the data in the intermediate regime of $H_T$.
Nevertheless, it reproduces well the global behavior: the signal
increases with $H_T$ until peak B first disappears, followed by peak
A. Using a semi-classical model (strictly valid for $S\gg1$), an
analytical expression for the linewidth of the EPR signal measured as
$H_z$ is swept can be obtained:
\begin{equation}
  \Delta H_z=\frac{hf\delta E}{(2g\mu_BS_z^0)^2H_z}
\end{equation}
where $\delta E$ is the energy broadening of the two levels and
$S_z^0=S\sqrt{1-(H_T/H_a)^2}$. The dashed line in the inset (b) of
Fig.~\ref{EPR} is a fit of the data with respect to $\delta E$. The fitting
value corresponds to a transverse relaxation time of 0.2~ns.  This
should be viewed as a lower bound of the decoherence time,
$T_2>0.2$~ns, which is consistent with ref.~\cite{barco04b}. We
emphasize that no photon induced changes in the magnetization have
been observed on this crystal, either at that low MW power, or in
higher power pulsed MW experiments. Using Eq.~\ref{relax}, one can
then deduce an upper bound for $T_1$ of $500~\mu$s, \emph{i.e.} much
less than it had been found in ref.~\cite{barco04b}.

\begin{figure}
  \includegraphics[width=6cm]{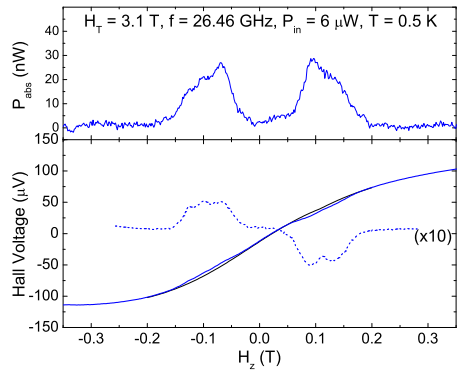}
  \caption{EPR signal (top) and photon-induced changes in the
    magnetization (bottom) of a Ni$_4$ crystal. The thin line is the
    equilibrium magnetization curve (MW off) and the dashed line the
    difference of magnetization curves with MW on and off expanded by
    a factor 10. The current in the Hall sensor was 10~$\mu$A.}
  \label{Mchange}
\end{figure}

Fig.~\ref{Mchange} presents the EPR signal and the simultaneous
measured changes in the magnetization of a different crystal than
reported in Fig.~\ref{EPR}. The same EPR spectroscopy study as that
described above yields similar results. However in that case, there is
a clear evidence of changes in the magnetization, at a similar level
of absorbed power. This is due to a much longer $T_1$, found to be
$30$~ms assuming that the same part of the crystal contributes to the
signals measured by the Hall sensor and the PNA. We note that using
the calibrated MW field strength, $h_{ac}=9.8\pm1.5$~mOe, the estimation
from the transition rate (Eq.~\ref{rate}) yields a power absorbed 30
times larger than that observed. One can thus infer that only 3\% of
the crystal is absorbing energy at the maximum of the resonance. We
attribute that to inhomogeneous broadening of the absorption line.

In conclusion, simultaneous measurements of magnetization and EPR
spectroscopy on tiny SMM single crystals have been demonstrated using
an integrated sensor. The obtained EPR data are well explained using a
direct diagonalization of the Hamiltonian of the spin system.
Associated photon-induced changes in the magnetization depend on the
the energy relaxation time. The latter is seen to vary greatly between
different crystals. This is probably linked to the thermalization of
the crystal \cite{garanin06}. A detailed study of such effects is in
progress. Finally, the integrated sensors developed for this work are
also promising to conduct Rabi and spin-echo experiments.  Such
experiments and real-time measurement of the magnetization should
allow measurements of the intrinsic decoherence time and relaxation
times in SMMs.

This work was suppported by NSF (Grant No. DMR-0506946).

\end{document}